\documentclass[aip,apl,preprint,floatfix]{revtex4-1}
\usepackage{graphicx}

\begin{document}
\title{Mechanisms of fragmentation of Al-W granular composites under dynamic loading}
\author{K.L. Olney}
\affiliation{Department of Mechanical and Aerospace Engineering, University of California at San Diego, La Jolla, California 92093-0411, USA}
\author{V.F. Nesterenko}
\affiliation{Department of Mechanical and Aerospace Engineering, University of California at San Diego, La Jolla, California 92093-0411, USA}
\affiliation{Materials Science and Engineering Program, University of California at San Diego, La Jolla, California 92093-0411, USA}
\author{D.J. Benson }
\affiliation{Department of Structural Engineering, University of California at San Diego, La Jolla, California 92093-0085, USA}
\date{\today}

\begin{abstract}
Numerical simulations of Aluminum (Al) and Tungsten (W) granular composite rings under various dynamic loading conditions caused by explosive loading were examined. Three competing mechanisms of fragmentation were observed: a continuum level mechanism generating large macrocracks described by the Grady-Kipp fragmentation mechanism, a mesoscale mechanism generating voids and microcracks near the unbonded Al/W interfaces due to tensile strains, and mesoscale jetting due to the development of large velocity gradients between the W particles and adjacent Al. These mesoscale mechanisms can be used to tailor the size of the fragments by selecting an appropriate initial mesostructure for a given loading condition.
\end{abstract}
\pacs{}

\maketitle

Granular materials are used to enhance the release of energy by mixing metal particles, e.g., Al, with traditional energetic materials~\cite{davis1} or by surrounding traditional energetic materials with a granular ``shell''.~\cite{frost,zhang2} The performance of these composite energetic materials is heavily influenced by the particle size of the metal powder which dictates the speed at which these particles can be oxidized by the detonation products or by the surrounding air.~\cite{davis1,frost,zhang2,beckstead} It has been shown that small sized suspended Al particles (40 microns) in air can sustain a detonation wave caused by the fast energy release due to the oxidization of these Al particles.~\cite{zhang3} Impact-initiated energetic materials  (metal powders in a polymer binder, consolidated powders, or mixtures of powders) are non-energetic under static or quasi-static loading but release energy under high strain, high strain rate deformations.~\cite{ames,mock,higgens1} In these materials, the material properties such as the mesostructure and the individual constituents' strength are instrumental in determining the performance characteristics.~\cite{ames} 

Another interesting class of reactive materials are granular composite materials that are able to carry a structural load under quasi-static loading combined with the ability to undergo a rapid bulk distributed pulverization under dynamic loading, releasing a large amount of usable chemical energy. The performance of these materials is determined by the properties of the individual components and the mesostructure. Al-W granular composites with an Al matrix comparable in strength to Al and have demonstrated an ability to undergo bulk distributed fragmentation under dynamic loading.~\cite{chiu1,chiu3} The post critical behavior of these Al-W granular composite materials was examined showing that the mesostructure of the W particles was the main factor influencing the fragment size.~\cite{olney1,nester2,olney2} 

For explosively driven homogeneous expanding rings, the mechanism of fragmentation is determined by the development of macrocracks~\cite{mott1} with a fragment size distribution that can be estimated using the Grady-Kipp equations.~\cite{grady} The typical fragment sizes for an explosively driven expanding ring made from Al 6061-T6 are of the order of 10mm.~\cite{nester2} Small scale experiments with explosive driven Al-W granular composite rings processed from -325 mesh Al and W particles by cold isostatic pressing (CIPing) were conducted in Ref.~\onlinecite{nester2}. The recovered fragments demonstrated that the mesostructure of the Al-W composite allowed for the generation of fragments with a size scale on the order of 100 microns. This reduction in the order of magnitude in fragment size suggested that there is a shift in the fragmentation mechanism from the continuum scale to that of the mesoscale determined by the mesostructure of the composite material. This mesoscale mechanism of fragmentation was attributed to the development of gradients in the particle velocities between the Al and the W constituents due to the significant difference between the shock impedances of Al and W.

In this paper, numerical simulations were performed to elucidate the shift in the mechanism from the continuum level determined by the nucleation, propagation and interaction of the macrocracks~\cite{grady} seen in the homogeneous samples to the generation of fragments which size is controlled by the mesoscale. The understanding of this new mechanism may allow for the tailoring of the size of fragments and the reactivity by selecting an appropriate initial mesostructure.

The two dimensional mesostructures in the numerical simulations were generated by randomly placing W particles such that the volume content of W in the composite was 30\%. This volume content corresponds to the mesostructures used in previous experiments for Al-W granular composites.~\cite{chiu1,chiu3,olney1,nester2,olney2} The shock pressure and temperature seen in the simulations performed in this paper reach conditions~\cite{nesterbook} that facilitate the bonding between the Al/Al interfaces during shock consolidation. However, the conditions attained during the shock loading do not cause bonding between the Al/W interfaces. As such, the Al particles are assumed to be fully bonded together creating a solid fully dense Al matrix. W particles are embedded into this matrix and are allowed to separate from the surrounding Al matrix. This mesostructure corresponds to a sample that has been processed by CIPing followed by hot isostatic pressing (HIPing) of the granular composite.~\cite{chiu1,chiu3,olney1,nester2,olney2}

A two-dimensional Eulerian hydrocode~\cite{benson} that allows for bonding/debonding~\cite{efrim1} was used to model the explosively driven granular composite ring expansion. Each material (Al, W, Cu, and the detonation products) in the sample used separate equations of state and mechanical properties. For Al, Cu, and W, the Johnson-Cook~\cite{johnson} material model with failure was used,
\begin{equation}
\sigma_{y}=[A+B\bar{\epsilon}_p^n][1+C\ln(\dot{\epsilon}/\dot{\epsilon_0})][1-T^{\star m}]
\end{equation}
\begin{equation}
\epsilon_f=[D_1 + D_2 exp{(D_3 \sigma^{\star})}][1+D_4 \ln(\dot{\epsilon}/\dot{\epsilon_0}][1+D_5 T^{\star}]
\end{equation}
\begin{equation}
\mathcal{D}=\sum{\frac{\Delta \epsilon_p}{\epsilon_f}}
\end{equation}
where $\bar{\epsilon}_p$ is the equivalent plastic strain, $\dot{\epsilon}$ is the strain rate,  $T^{\star}$ is the homologous temperature, $\sigma^{\star}$ is the pressure divided by the equivalent deviatoric stress, $\mathcal{D}$ is the damage parameter where the material is considered fully damaged and unable to support shear when this parameter reaches 1. $A, B, n, C, \dot{\epsilon_0}, m , D_1, D_2, D_3, D_4,$ and $D_5$ are parameters taken from the open literature.~\cite{johnson,holm,stein} This material model was used in conjunction with the Mie-Gr\"{u}neisen equation of state. The explosive was assumed to have an instantaneous detonation converting the explosive into detonation products. We used this approximation because the focus of our research was on the mechanism of fragmentation of the composite material under similar conditions of loading. The detonation products were modeled using the Chapman-Jouguet relations for the stationary detonation: 
\begin{equation}\label{cj1}
\rho_{CJ}=\rho_{0}\frac{\gamma_{CJ}+1}{\gamma_{CJ}}
\end{equation}
\begin{equation}\label{cj2}
P_{CJ}=\rho_0 \frac{D_{CJ}^2}{\gamma_{CJ}+1}
\end{equation}
where $\rho_{CJ}$, $\gamma_{CJ}$, $P_{CJ}$, and $D_{CJ}$ are the density, Gr\"{u}neisen parameter, pressure, and detonation velocity at the Chapman-Jouguet point.  

The detonation products were modeled using an ideal gas model:
\begin{equation}
P=(\gamma_{CJ} - 1)\frac{\rho}{\rho_{CJ}}E
\end{equation}
with an initial pressure corresponding to the Chapman-Jouguet detonation pressure. Parameters used in this model were calibrated by simulating an explosively driven expanding Cu ring and comparing the free surface velocity with experiments.

\begin{figure}[t]
\includegraphics{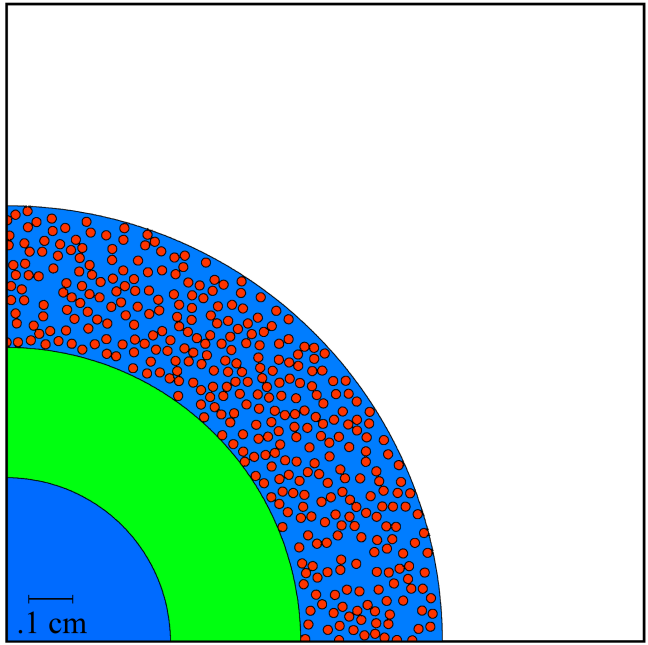}
\caption{\label{fig:initial_geo} (color online) Initial mesostructure for the CIPed+HIPed Al-W granular composite ring with a Cu inner liner. The central blue area represents the detonation products, the second layer (green) is the Cu inner liner, and the outer layer is the granular composite composed of 200 $\mu m$ diameter W particles (red) embedded in a solid Al matrix (blue). The W particles were randomly placed such that the volume content of W in the composite was 30\%. The mesostructure shown for the Al-W composite was used for all of the simulations presented in this paper. For the simulations where the detonation products are in direct contact with the composite ring, the inner Cu liner was removed and replaced with the detonation products.}
\end{figure}

The initial mesostructure of the Al-W composite used in the simulations is presented in Fig.~\ref{fig:initial_geo}. In this paper, four variations of this initial setup were examined to elucidate the influence shock amplitude on the mechanisms of fragmentation. The calculations were performed at two initial pressures of detonation products with parameters corresponding to Primasheet 1000 (a ``weak'' explosive, used in experiments in Ref.~\onlinecite{nester2}): $D_{CJ}$=0.68 $cm/\mu s$ $\gamma_{CJ}$=3 $\rho_{0}$=1.46 $g/cm^3$ resulting in $P_1$=0.168 Mbar and  LX-14 (a ``strong'' explosive): $D_{CJ}$=0.88 $cm/\mu s$ $\gamma_{CJ}$=2.84 $\rho_{0}$= 1.835 $g/cm^3$ resulting in $P_2$=0.37 Mbar. The influence of the Cu inner liner on the mechanisms of fragmentation was also examined.

\begin{figure}[t]
\includegraphics{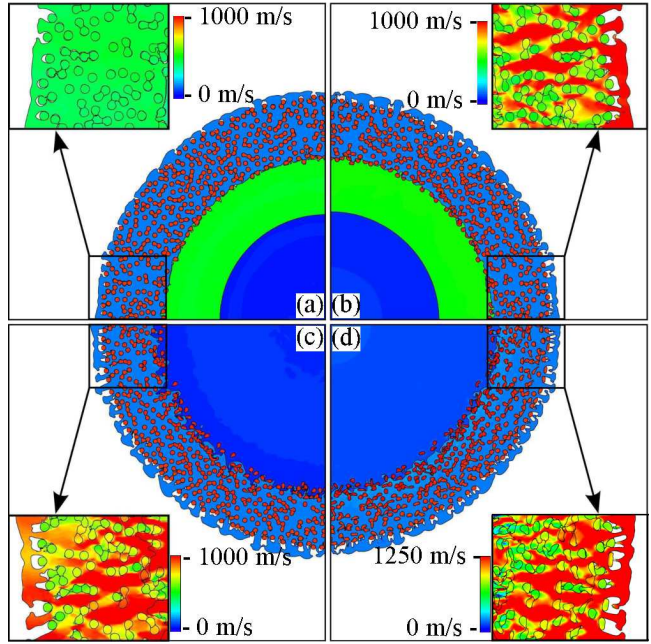}
\caption{\label{fig:shock_wave1} (color online) Patterns of mesoscale fragmentation after the granular composite ring has radially expanded by 10\% for different initial geometries and detonation product pressures: (a) Case 1, $P_{CJ}$=0.168 Mbar  with a Cu inner liner, (b) Case 2, $P_{CJ}$=0.37 Mbar with a Cu inner liner, (c) Case 3, $P_{CJ}$=0.168 Mbar with detonation products in direct contact with the granular composite, and (d) Case 4, $P_{CJ}$=0.37 Mbar with detonation products in direct contact with the granular composite. Detailed views of the radial velocity in the granular composites are plotted to show the velocity gradients between W particles and the surrounding Al. These velocity gradients are a driving force for mesoscale fragmentation at later stages of expansion.}
\end{figure}
\begin{figure}[t]
\includegraphics{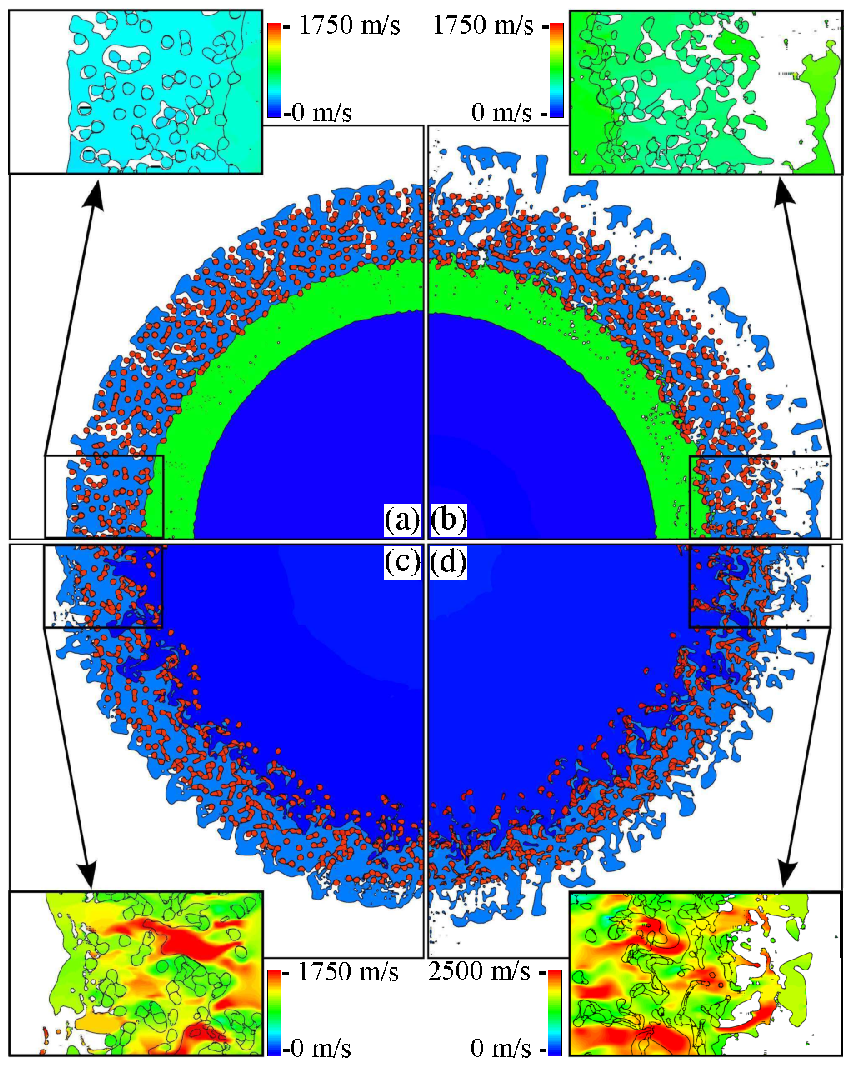}
\caption{\label{fig:shock_wave2} (color online) Patterns of fragmentation in the granular composite rings at 50\% radial expansion. A previous stage of expansion was presented in Fig.~\ref{fig:shock_wave1} (a), (b), (c), and (d) corresponding to Cases 1-4 respectively. Detailed views of the radial velocity in the granular composite are plotted to show the equilibration of the velocities between the Al and W constituents at this stage of expansion.}
\end{figure}

Results of the numerical simulations are presented in Fig.~\ref{fig:shock_wave1} and Fig.~\ref{fig:shock_wave2}. These figures show the composite in all four configurations with the two variations in detonation products corresponding to Case 1 with detonation pressure of $P_1$=0.168 Mbar and a Cu inner liner, Case 2 with a detonation pressure of $P_2$=0.37 Mbar with a inner Cu liner, and Case 3 and Case 4 that have the same detonation pressures as Case 1 and Case 2 respectively but without the Cu inner liner. In Case 3 and Case 4, the Cu inner liner was replaced with the detonation products such that the detonation products directly contacted the Al-W granular composite. Due to the variations in the initial detonation pressures and differences in the subsequent expansion rates, all samples depicted in Fig.~\ref{fig:shock_wave1} and Fig.~\ref{fig:shock_wave2} are compared at the same radial expansion (10\% and 50\% increase in the initial radius). 

At 10\% radial expansion, velocity gradients between the lighter Al matrix and the heavier W particles have formed due to the initial shock loading. The magnitude of the particle velocity gradient shows a strong dependency on the amplitude of the shock loading. This can be clearly seen when comparing Case 1 and Case 2: the velocity gradients between the Al and W components are negligible in Case 1 while they are significant in Case 2. In Cases 2-4, the velocity gradients between the Al and W were large enough to cause microjetting to occur within the composite and the formation of microjets near the vicintity of the free surface. The addition of the Cu inner liner significantly reduced the magnitude of the velocity gradients by reducing the amplitude of the initial shock wave. This is especially dramatic when comparing Case 1 with Case 3 where the addition of the Cu inner liner reduced the velocity gradients between the Al and W to levels where microjetting did not occur. In Case 2, the added Cu inner liner reduces the magnitude of the velocity gradients by about  20\% in comparison to Case 4. This suggests that the addition of the Cu inner liner reduces the effectiveness of the microjetting mechanism and in the cases with weak detonation pressures, causes this mechanism to be inactive. In the Case 2 and Case 4, these microjets eject approximately 30\% of the Al from the composite while no Al ejection was seen in Case 1 or Case 3. 

As the ring expands in all cases, the velocity gradients begin to diminish due to shock wave equilibration and dissipation. The rate at which this equilibration occurs is largely dependent on the shock amplitude and the shock impedances of the components. As the particle velocity gradients diminish, the mesoscale microjets cease to exist.  This results in a shift in the mechanism of fragmentation from the ejection of Al due to gradients in the particle velocity to the competition between the continuum scale development of macrocracks and the networks of microcracks developing on the mesoscale between neighboring W particles.

Figure~\ref{fig:shock_wave2} depicts the samples after expanding 50\% of the initial radius. It is clear from this figure that the differences in the detonation products create qualitatively different fracture patterns in the composite material. In Case 1, potential fragments are forming with the same size scale as those predicted by the Grady-Kipp equations. In Case 2, the composite shows a large number of mesoscale voids opening around the W particles and their coalescence with nearby voids. The W particles in these regions exhibit a tendency to form clumps of 5-10 particles. This suggests that a subsequent fragments will be generated on a smaller size scale containing about 5-10 W particles.   

In the simulations without the Cu inner liner, the W particles were heavily deformed. Furthermore, in the regions of the composite near the detonation product interface, the Al matrix was stripped from the W particles leaving free W partilces behind. In these cases, the detonation products penetrate into the composite where the Al material is fully damaged, resulting in a Rayleigh-Taylor type instability. These instabilities grow into the composite, channeled by clumps or short chains of W particles, creating ``fingers'' of detonation products that introduce a new size scale of fragmentation based on the mesoscale. The paths of these ``fingers'' start  in the same location on the inner Al-W composite ring but the penetration depth at a given radial expansion is determined by the detonation pressure.

The amplitude of the initial shock wave, dictated by the pressure of the detonation products, determines the mechanism of pulverization for the Al in the granular composite. At low detontion pressures, generating relatively low amplitude shock waves, the disintegration is mainly determined by the competition between the continuum Grady-Kipp mechanism of macrocracking and the opening of mesoscale voids due to the Al/W interfaces that are initially not bonded due to local tensile strains when the composite radially expands. At higher detonation pressures that generate larger amplitude shock waves, a third fragmentation mechanism develops; the mesoscale jetting due to the large gradient of particle velocities between the W and Al in the regions adjacent to the free surface. This mechanism is responsible for the pulverization of approximately of 30\% of Al. The remainder of the composite is pulverized at later stages of expansion when the gradients in the velocity between the Al and the W vanish due to wave equilibration. This pulverization is based on the competition between the mechanisms of macrocracking and the opening of the mesoscale voids at the Al/W interfaces. In the cases where the Cu inner liner was not present, the detonation products penetrated into the composite generating an additional mesoscale mechanism of fragmentation based on the mesostructure. These mesoscale mechanisms can be used to tailor the size of the fragments by selecting the appropriate initial mesostructure. At relatively low shock pressures, the initially unbonded interfaces between particles may be the most important mesostructural factor governing the size of the fragments while at higher detonation pressures, the mesoscale jetting can be the most important factor for pulverization.
\section*{ACKNOWLEDMENTS}
The support for this project provided by the Office of Naval Research Multidisciplinary University Research Initiative Award N00014-07-1-0740, program manager Dr. Clifford D. Bedford.

\bibliography{draft1}

\end{document}